%% file: micro_edf.tex
\documentclass[a4paper]{jpconf}
\usepackage[utf8]{inputenc}
\usepackage{graphicx}
\usepackage{lmodern}
\usepackage{amsmath}
\usepackage{bm}

\usepackage[square,sort&compress,numbers]{natbib}
\usepackage[dvipsnames]{xcolor}
\usepackage[colorlinks=true, urlcolor=Brown, linkcolor=RoyalBlue,
            citecolor=Green]{hyperref}
\usepackage{glossaries-extra}
\include{acronyms}

\begin{document}

\hfill LLNL-PROC-769943

\title{Microscopically based energy density functionals for nuclei using the density matrix expansion}

\author{R. Navarro Pérez$^{1,2}$, N. Schunck$^2$}

\address{$^1$ Department of Physics, San Diego State University, San Diego, California 92182, USA}
\address{$^2$ Nuclear and Chemical Sciences Division, Lawrence Livermore National Laboratory, Livermore, CA 94551, USA}

\ead{rnavarroperez@sdsu.edu}

\begin{abstract}
While ab initio many-body techniques have been able to successfully describe the properties of light and intermediate mass nuclei based on chiral effective field theory interactions, neutron-rich nuclei still remain out of reach for these methods. Conversely, energy density functional approaches can be used to calculate properties of heavy nuclei but rely mostly on phenomenological interactions. A usable form of the nuclear energy density functional that is rooted in the modern theory of nuclear forces was presented recently. The first component of this new set of functionals corresponds to the direct part (Hartree term) of the expectation value of local chiral potentials on a Slater determinant. The exchange term, which is a functional of the non-local density, is transformed into a local functional by applying the density matrix expansion. In order to reduce the computational cost due to the direct implementation of non-separable, local interactions in the Hartree term, we use an approximation to represent the regularized Yukawa functions in terms of a sum of (separable) Gaussian functions. These proceedings analyze the accuracy of such an approximation in terms of the number of Gaussian functions and look for an optimal value that gives an acceptable level of accuracy while maintaining the computational memory requirements in a many-body calculation as low as possible.
\end{abstract}

\section{Introduction}

One of the overarching goals of nuclear physics is to find a theory capable of describing nuclear structure and reactions in terms of the underlying interactions between quarks and gluons. The description of the interaction between nucleons in \gls{CEFT} provides the desired microscopic underpinning and also exhibits systematic, order-by-order, improvement~\cite{furnstahl_chiral_1997, bedaque_low_2003, ekstrom_optimized_2013, valderrama_power_2015}. Ab initio methods employ these realistic interactions and calculate properties of nuclear systems by directly considering the interaction between each pair of nucleons. This approach has been successful in describing light and medium nuclei (up to $Z=50$)~\cite{hergert_nuclear_2018}. The same systematic, order-by-order, improvement from the interaction can be observed in the description of nuclear properties calculated with ab initio methods~\cite{hagen_emergent_2016}. This allows keeping track of the predictive power of these type of theoretical calculations when looking at properties that have not been measured yet. However heavy systems, particularly neutron-rich systems, remain out of reach for ab initio methods. These neutron-rich systems are of great interest in view of upcoming measurements in flagship facilities like \gls{FRIB} and for the description of astrophysical phenomena like the r-process~\cite{engel__1999, surman_nucleosynthesis_2006}.

In contrast to ab initio methods, \gls{EDF} methods, also known as self-consistent mean-field approaches, substitute the interaction  due to all pairs of nucleons by an ``average'' of the interaction with all the other nucleons \cite{schunck2019,bender2003}. This substitution significantly reduces the computational cost of many-body nuclear calculations and allows estimating properties of heavy systems such as the neutron-rich nuclei mentioned earlier. However, the form of the interaction used in \gls{EDF} approaches is usually constructed \emph{ad hoc} in order to simplify calculations, and parameters are adjusted to reproduce experimental nuclear properties such as masses, charge radii or spin-orbit splittings. This leads to effective interactions of rather phenomenological nature whose predictive power, when extrapolated to properties beyond the ones used to determine its parameters, can be hard to maintain or even estimate.

Following earlier studies \cite{stoitsov2010}, we have recently proposed a new method to determine an \gls{EDF} from microscopic nuclear potentials derived from \gls{CEFT} through a tool known as the \gls{DME}. In practice, the new \gls{EDF} looks like a phenomenological Skyrme-like contact interaction to which is added a more microscopic component which takes the form of density-dependent couplings resulting from applying the \gls{DME} to the non-local one-body density \cite{navarro_perez_microscopically_2018}. This approach encodes the long-range physics contained in the \gls{CEFT} potential. The phenomenological Skyrme-like part accounts for the short-range physics and its parameters need to be readjusted in order to reproduce experimental nuclear structure data. 

The computational implementation of these semi-phenomenological interactions involves using certain numerical approximations (beyond the \gls{DME} itself). In particular, the contribution of the long-range part of the chiral potential to the nuclear binding energy can be efficiently computed by approximating the Yukawa potentials with a sum of Gaussian functions. This approximation allows to recast the \gls{CEFT}-derived potential in a form analogous to the purely phenomenological Gogny interaction, taking advantage of fact that this type of interaction is already implemented in several many-body \gls{DFT} codes. Of course the implementation of any approximation requires to ensure a desired level of accuracy in order to avoid the introduction and propagation of numerical errors in all kinds of calculations. These proceedings analyze the convergence of the sum of Gaussian functions approximation to the central terms in the chiral potential at \gls{NLO} and \gls{N2LO}. Section \ref{sec:EDF} gives a brief overview of the implementation of semi-phenomenological \gls{EDF} introduced in~\cite{navarro_perez_microscopically_2018}. Following that, section \ref{sec:accuracy} presents an analysis of the accuracy of the Gaussian expansion depending on the number of terms included. Conclusions and outlook are discussed in section \ref{sec:conlcusions}.

\section{Semi-phenomenological EDF}
\label{sec:EDF}

Our many-body theoretical framework is the \gls{HFB} formalism. In this framework, the many-body wave function is taken as a quasiparticle vacuum \cite{schunck2019}. The total energy is expressed as a functional of the generalized density, or equivalently, as a functional of the one-body density matrix $\rho$ and the pairing tensor $\kappa$. The full \gls{EDF} can often be broken down into a sum of several terms, one that depends exclusively on $\rho$ (the particle-hole channel) and another one that also involves $\kappa$ and is only active when pairing correlations are present (the particle-particle channel).

The \gls{EDF} in the particle-hole channel is often constructed by representing the many-body wave function by a Slater determinant and using the Wick theorem to express the energy due to many-body interactions as a functional of the density of particles in the system. In the simplest case of two-body interactions, this leads to writing 
\begin{equation}
	E^{\rm NN}[\rho] = \frac{1}{2} \sum_{ij}\langle i j \left| V^{\rm NN} (1 - P_\sigma P_\tau P_r) \right| k l \rangle \rho_{ki} \rho_{lj},
\end{equation}
with $\rho_{ij}$  the matrix elements of the one-body density matrix on an arbitrary basis of the single-particle Hilbert space, $P_\sigma$, $P_\tau$ and $P_r$ the usual spin-, isospin- and position-exchange operators and $V^{\rm NN}$ the two-body nuclear potential. The anti-symmetrization operator $(1 - P_\sigma P_\tau P_r)$ results in direct and exchange contributions, also known as the Hartree and Fock energies. After transforming the spatial coordinates to relative~$(\boldsymbol{r})$  and center-of-mass~$(\boldsymbol{R})$ coordinates, the Hartree and Fock energies become
\begin{align}
	\label{eq:E_hartree}
 	E_{\rm Hartree}^{\rm NN} &= \frac{1}{2} \rm{Tr}_1 \rm{Tr}_2 \int d \boldsymbol{R}  \int d \boldsymbol{r} \langle V^{\rm NN}(\boldsymbol{r}) \rangle  
 	\rho_1\left(\boldsymbol{R} + \frac{\boldsymbol r}{2}\right) 
 	\rho_2\left(\boldsymbol{R} - \frac{\boldsymbol r}{2}\right) \\
 	\label{eq:E_fock}
 	E_{\rm Fock}^{\rm NN} &= -\frac{1}{2} \rm{Tr}_1 \rm{Tr}_2 \int d \boldsymbol{R}  \int d \boldsymbol{r} \langle V^{\rm NN}(\boldsymbol{r}) \rangle  
 	\rho_1\left(\boldsymbol{R} - \frac{\boldsymbol r}{2}, \boldsymbol{R} + \frac{\boldsymbol r}{2}\right) 
 	\rho_2\left(\boldsymbol{R} + \frac{\boldsymbol r}{2}, \boldsymbol{R} - \frac{\boldsymbol r}{2}\right) P_{12}^{\sigma \tau},
\end{align}
where the potential bracket includes spin and isospin states, the traces refer to summations over those quantum numbers and the local density matrix is $\rho(\boldsymbol{x}) = \rho(\boldsymbol{x},\boldsymbol{x})$; for details on how to incorporate three-body forces into this formalism, see refs.~\cite{dyhdalo_applying_2017,navarro_perez_microscopically_2018}. In a purely phenomenological approach the spatial form of the two-body interaction $V^{\rm NN}$ is partly chosen to simplify the calculation of these energy terms. Popular choices include delta functions for Skyrme functionals~\cite{skyrme1959,stone2007} and Gaussian functions for the Gogny interaction~\cite{decharge1980}. Note that by representing the many-body wave function with a Slater determinant, the \gls{EDF} approach assumes independent particles. Many-body correlations are recovered by a combination of two mechanisms: the spontaneous breaking of symmetries of the nuclear potential and the calibration of the parameters of the energy-density dependent functional to experimental data \cite{schunck2019}.

In order to anchor the \gls{EDF} approach to a more microscopic theory and a systematic, order-by-order, description of nuclear forces, the new microscopic functionals presented in~\cite{navarro_perez_microscopically_2018} start from the local chiral potentials up to \gls{N2LO} with and without $\Delta$ excitations as outlined in~\cite{piarulli_minimally_2015, piarulli_local_2016}. A direct substitution of these potentials in equations (\ref{eq:E_hartree}) and (\ref{eq:E_fock}) instead of the purely phenomenological Gogny or Skyrme interactions results in computationally more expensive numerical integrals that would need to be calculated at every \gls{HFB} iteration. To sidestep the increase in computational cost two approximations are made. The spatial form factor of the potential in equation (\ref{eq:E_hartree}) is approximated by a sum of Gaussian functions whose strength and range parameters are adjusted to reproduce the original chiral interaction. The use of Gaussian functions allows treating the Hartree term in the same fashion as a Gogny interaction and make use of previously benchmarked many-body \gls{DFT} codes in which Gogny interactions have already been implemented, such as \texttt{HFBTHOv3}~\cite{navarro_perez_axially_2017}. The second approximation is to employ the \gls{DME} formalism to calculate the Fock energy term. The \gls{DME} expands the non-local density in equation (\ref{eq:E_fock}) in terms of a local density and derivatives of that local density as
\begin{equation}
	\rho\left(\boldsymbol{R} + \frac{\boldsymbol r}{2}, \boldsymbol{R} - \frac{\boldsymbol r}{2}\right) \approx \Pi_0^\rho(k_F r)\rho(\boldsymbol{R}) + \frac{r^2}{6}\Pi_2^\rho(k_F r) \left[\frac{1}{4} \Delta \rho(\boldsymbol{R}) - \tau(\boldsymbol{R})  + \frac{3}{5} k_F^2 \rho(\boldsymbol{R}) \right],
\end{equation}
where the $\Pi$ functions are proportional to a reduced spherical Bessel function, $k_F$ is the usual Fermi momentum and $\tau(\boldsymbol{r})$ is the kinetic density given by
\begin{equation}
	\tau(\boldsymbol{r}) = \nabla \cdot \nabla' \rho(\boldsymbol{r},\boldsymbol{r}')|_{\boldsymbol{r}=\boldsymbol{r}'}.
\end{equation}
As a result, the \gls{DME} turns the finite-range part of the local, \gls{CEFT} chiral potential into a Skyrme-like functional, the main difference with traditional Skyrme functionals being that all coupling constants are now functionals of the local density. 

The procedure briefly outlined above results in a semi-phenomenological approach in which the re-parametrization of the Skyrme-like part encodes the zero-range, i.e., high-energy, part of the chiral potential and represents the short-range interaction, while the finite-range part of the interaction provides the order-by-order improvement to the description of many-body potentials. With this approach it is in principle possible to construct a scalable framework capable of describing neutron-rich nuclei with a series of systematic improvements.

\section{Numerical accuracy of the approximation for the Hartree term}
\label{sec:accuracy}

As mentioned in the previous section, the implementation of the microscopically constrained density functionals requires approximating the chiral potential in equation (\ref{eq:E_hartree}) as a sum of Gaussian functions,
\begin{equation}
	\label{eq:SumOfGaussians}
	V_\chi (r) \rightarrow \tilde{V}(r) = \sum_{i=1}^N V_i e^{-r^2/\mu_i}
\end{equation}
It is crucial then to ensure that this Gaussian expansion gives a correct description of the actual chiral interaction in terms of Yukawa potentials.

An important feature of the local chiral potentials of \cite{piarulli_minimally_2015, piarulli_local_2016} is that the potential contains a regulator in order to avoid divergences at short distances. The regulator has the form
\begin{equation}
	f(r) = \left[1 - \exp\left(- \frac{r^2}{R_c^2} \right) \right]^n.
\end{equation}
The potentials implemented in~\cite{navarro_perez_microscopically_2018} in particular use $R_c = 1.0$ fm and $n=6$. This regulator ensures that the interaction vanishes as $r \rightarrow 0$. A simple way to reproduce this kind of behavior in the sum of Gaussian functions in equation (\ref{eq:SumOfGaussians}) is to impose the condition,
\begin{equation}
	\label{eq:GaussConstraint}
	V_N = - \sum_{i=1}^{N-1} V_i.
\end{equation}

The choice for the number of Gaussian functions used in the approximation is dictated by the necessary compromise between precision and available resources. Using too few Gaussian functions causes a poor approximation that will introduce numerical errors in the \gls{HFB} many-body calculation; using too many Gaussian functions significantly increases the necessary memory to store all non-zero one-body density matrix elements with the possibility of going beyond the amount of memory available in current high performance computing architectures. With these restrictions in mind it is worth looking at the difference between the actual chiral interaction and equation (\ref{eq:SumOfGaussians}) as a function of the number of Gaussian terms included in the approximation. While these proceedings focus in particular on the $W_C^{\rm NLO}$ and $V_C^{\rm N2LO}$ terms in the chiral interaction outlined in~\cite{piarulli_minimally_2015}, the analysis presented here applies as well to the other central terms at different orders with and without the inclusion of $\Delta$ excitations.

\begin{figure}[!ht]
\begin{center}
\includegraphics[width = 0.9\linewidth]{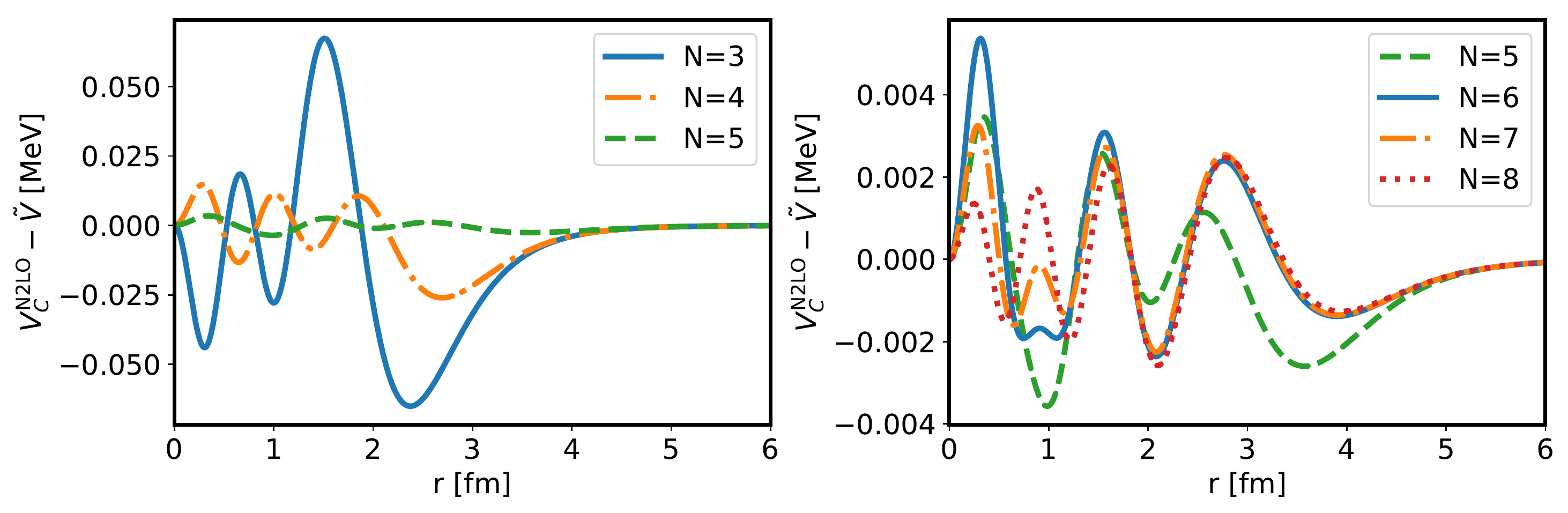}
\end{center}
\caption{\label{fig:Vc_diff}Difference between the regularized central term $V_C$ in the local chiral potential at N2LO as described in~\cite{piarulli_minimally_2015} and the Gaussian expansion of equation (\ref{eq:SumOfGaussians}) with different number of Gaussian functions $N$. The $N = 5$ case is included in both panels for comparison}
\end{figure}

\begin{figure}[!ht]
\begin{center}
\includegraphics[width = 0.9\linewidth]{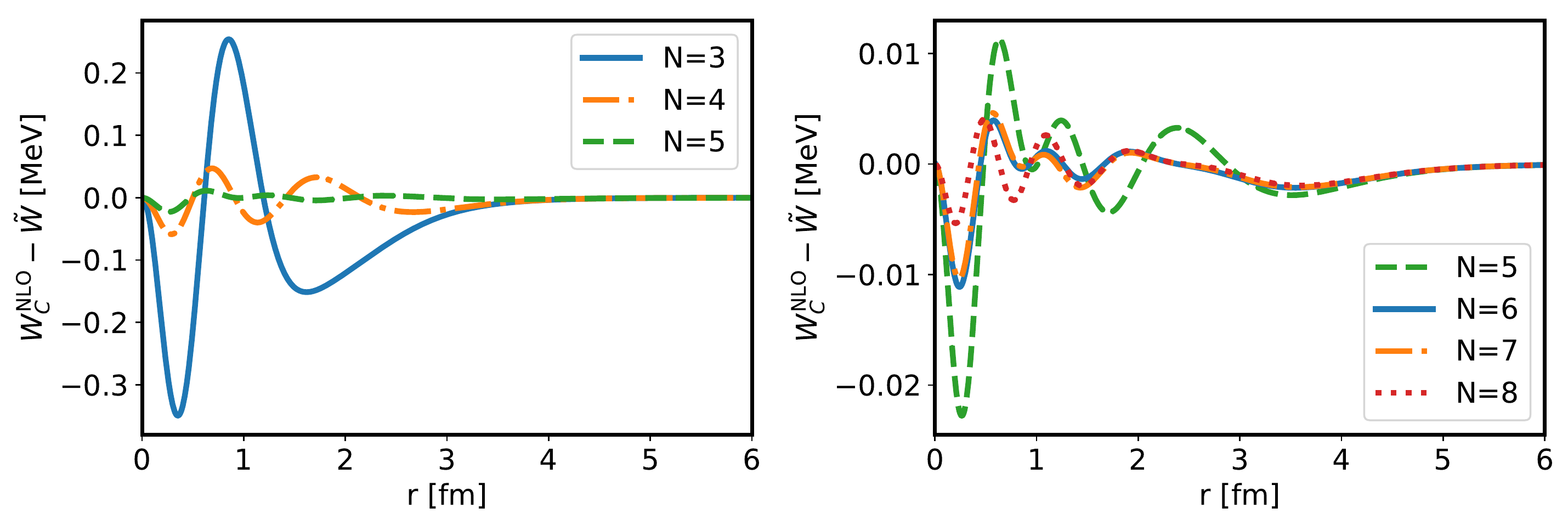}
\end{center}
\caption{\label{fig:Wc_diff} Same as figure \ref{fig:Vc_diff} for the central isospin-dependent $W_C$ term at NLO}
\end{figure}

For each of the two interactions mentioned above, we study six different approximations that go from using only three Gaussian functions up to a total of eight. In each approximation the $V_i$ and $\mu_i$ parameters are adjusted, imposing the constraint of equation (\ref{eq:GaussConstraint}), to reproduce the chiral interaction as well as possible. While a regularized approximation ($\tilde{V}\rightarrow 0$ as $r \rightarrow 0$) can already be constructed using only two Gaussian functions, a reasonable approximation is obtained starting at $N=3$.
The same type of analysis for the central isospin term $W_C^{\rm NLO}$ is presented in figure \ref{fig:Wc_diff}. Although the curves in the right panel show a reduction in the difference between the chiral potential and the sum of Gaussian functions approximation from $N=5$ to $N=6$, the order for magnitude remains the same all the way up to $N=8$. Considering that the computational memory requirements would be considerably increased by an extra Gaussian function it is preferred to keep the approximation at $N=5$ as the optimal compromise between numerical accuracy and memory requirements. 

Although the figures \ref{fig:Vc_diff} and \ref{fig:Wc_diff} give visual evidence that increasing the number of Gaussian functions beyond $N=5$ adds little to no improvement, it is preferred to quantify this statement via a numerical calculation. To estimate the average discrepancy between the chiral interaction and the Gaussian approximation \eqref{eq:SumOfGaussians}, we calculate the root mean square deviation using a fine grid sampling in radius every $0.01$ fm from $r=0$ fm to $r=10$ fm. The numerical results are shown in table \ref{tab:rms} and quantifies what was already suggested in the previous figures; there is little to no change in the error beyond $N=5$.

\begin{table}[!ht]
\caption{\label{tab:rms} Root mean square deviation between the two central terms $V_C$ and $W_C$ described in~\cite{piarulli_minimally_2015} and the Gaussian expansion in equation (\ref{eq:SumOfGaussians}) for different number of Gaussian functions.}
\begin{center}
\begin{tabular}{lll}
\br
$N$ & $V_c$ & $W_c$ \\
 & MeV & MeV \\
\mr
3&0.0229&0.0828\\
4&0.0083&0.0155\\
5&0.0013&0.0039\\
6&0.0013&0.0018\\
7&0.0010&0.0017\\
8&0.0010&0.0012\\
\br
\end{tabular}
\end{center}
\end{table}

\section{Conclusions and outlook}
\label{sec:conlcusions}

The implementation of microscopically-constrained \gls{EDF} allows performing many-body calculations of large neutron-rich systems with a systematic order-by-order improvement determined by \gls{CEFT}. The performance of this new family of functionals has shown an improved description of nuclear masses when compared to other state-of-the-art functionals like UNEDF2~\cite{navarro_perez_microscopically_2018}. This milestone paves the way towards performing calculations relevant to astrophysics, such as simulations of the r-process, with nuclear models with a firm microscopic underpinning stemming from quarks and gluons. Given the potential of these new functionals, it is important to verify that the approximations employed for the implementation have a reasonable level of accuracy in order to avoid compromising the numerical precision of upcoming calculations. These proceedings focused on the numerical accuracy of the sum of Gaussian functions used to represent the central terms of the chiral potential in the calculation of the Hartree energy term. While an infinite number of Gaussian functions could potentially guarantee numerically exact representation of the regularized Yukawa potentials in the chiral interaction, computational resources are limited and a large number of Gaussian functions results in a prohibitively large amount of computational memory. This problem can be mitigated by finding the smallest number of Gaussian functions in the approximation that results in an acceptable level of accuracy. 

While these proceedings focused on the sum of Gaussian functions approximation, the numerical accuracy of other approximations and numerical integrals was verified during the implementation of the new family of microscopic functionals. However, other sources of uncertainty remain present and need to be quantified and propagated. For example, beyond mean-field effects near closed shell nuclei, which are not described within the single-reference \gls{EDF} approach used here, can clearly be seen in figure 5 of~\cite{navarro_perez_microscopically_2018}. This shows a clear and systematic discrepancy between experimental values and the \gls{HFB} approach and tools for modeling this discrepancy need to be implemented. An approach based on Bayesian Neural Networks has been proposed for this very particular type of discrepancy in simpler mass models like the liquid drop model~\cite{utama_nuclear_2016}, and a similar approach can be directly applied to density based calculations within the \gls{HFB} formalism.

\section{Acknowledgments}
Support for this work was provided through the Scientific Discovery through Advanced Computing (SciDAC) program funded by U.S. Department of Energy, Office of Science, Advanced Scientific Computing Research and Nuclear Physics. It was partly performed under the auspices of the US Department of Energy by the Lawrence Livermore National Laboratory under Contract DE-AC52-07NA27344. Computing support for this work came from the Lawrence Livermore National Laboratory (LLNL) Institutional Computing Grand Challenge program.

\bibliography{biblio,zotero}

\end{document}

%% file: acronyms.tex
\setabbreviationstyle[acronym]{long-short}

\glssetcategoryattribute{acronym}{nohyperfirst}{true}

\newacronym{CEFT}{$\chi$-EFT}{Chiral Effective Field Theory}
\newacronym{LECs}{LECs}{Low Energy Constants}
\newacronym{FRIB}{FRIB}{the Facility for Rare Isotope Beams}
\newacronym{DFT}{DFT}{Density Functional Theory}
\newacronym{PI}{PI}{Principal Investigator}
\newacronym{HPC}{HPC}{High Performance Computing}
\newacronym{DME}{DME}{Density Matrix Expansion}
\newacronym{UQ}{UQ}{Uncertainty Quantification}
\newacronym{NN}{NN}{Nucleon-Nucleon}
\newacronym{NSAC}{NSAC}{Nuclear Science Advisory Committee}
\newacronym{MCMC}{MCMC}{Markov Chain Monte Carlo}
\newacronym{EDF}{EDF}{Energy Density Functional}
\newacronym{NLO}{NLO}{next to leading order}
\newacronym{N2LO}{N2LO}{next to next to leading order}
\newacronym{HFB}{HFB}{Hartree-Fock-Bogoliubov}